\documentclass[amsmath,amssymb,prl,superscriptaddress,preprint]{revtex4-1}

\usepackage{amsmath,amssymb,amsfonts,mathrsfs}
\usepackage{graphicx,soul}
\usepackage[ansinew]{inputenc}
\usepackage{color,ulem}
\usepackage{ulem}

\newcommand {\peter}[1]{{\color{black} #1}}

\newcommand {\jw}[1]{{\color{black} #1}}
\newcommand {\jwb}[1]{{\color{black} #1}}
\newcommand {\jwc}[1]{{\color{black} #1}}

\begin{document}

\title{\jwc{Tunneling into} thin superconducting films: interface-induced \jwb{quasiparticle} lifetime reduction}

\author{P.~L\"{o}ptien}
\affiliation{Institute of Applied Physics, University of Hamburg, D-20355 Hamburg, Germany}
\author{L.~Zhou}
\email[Electronic mail: ]{L.Zhou@fkf.mpg.de}
\thanks{\jwc{Present address: Max Planck Institute for Solid State Research, D-70569 Stuttgart, Germany}}
\affiliation{Institute of Applied Physics, University of Hamburg, D-20355 Hamburg, Germany}
\author{A.~A.~Khajetoorians}
\thanks{\jwc{Present address: Institute for Molecules and Materials (IMM), Radboud University, 6525 AJ Nijmegen, The Netherlands.}}
\affiliation{Institute of Applied Physics, University of Hamburg, D-20355 Hamburg, Germany}
\author{J.~Wiebe}
\affiliation{Institute of Applied Physics, University of Hamburg, D-20355 Hamburg, Germany}
\author{R.~Wiesendanger}
\affiliation{Institute of Applied Physics, University of Hamburg, D-20355 Hamburg, Germany}

\date{\today}

\begin{abstract}
Scanning tunneling spectroscopy \jwc{measurements} of superconducting thin lanthanum films grown on a normal metal tungsten substrate reveal an extraordinarily large \jwb{broadening} of the coherence peaks. The observed broadening corresponds to very short electron-like quasiparticle lifetimes in the tunneling process. A thorough analysis considering the different relaxation processes reveals that the dominant mechanism is an efficient \jwb{quasiparticle} relaxation at the interface between the superconducting film and the underlying substrate. This process is of general relevance to scanning tunneling spectroscopy studies on thin superconducting films \jwb{and enables measurements of film thicknesses via a spectroscopic method.}
\end{abstract}



\maketitle
In tunneling experiments with superconducting electrodes, the differential conductance $\mathrm dI/\mathrm dV$ resembles the superconducting density of states (DOS). 
In early studies, planar tunnel junctions with oxide layers \cite{Miller1967, Blackford1969, Dynes1978, Dynes1984, Lee1988}
or point contacts \cite{Grajcar1995}
were utilized to determine superconducting properties like the energy gap $\Delta$. With the ability to perform low temperature scanning tunneling spectroscopy (STS) in ultra-high vacuum (UHV), it has become possible to probe \textit{in situ} fabricated superconducting thin films with unprecedented control over the properties of the tunneling barrier interface, and to determine $\Delta$ with atomic-scale spatial resolution \cite{Rodrigo2004a, Nishio2006, Bergeal2006, Bose2009, Brihuega2011, Loeptien2014}.

In such experiments, $\mathrm dI/\mathrm dV$ is broadened at the coherence peaks if the lifetime $\tau$ of the quasiparticle states the electrons tunnel into is on the order of $\hbar/\Delta$. Such lifetime effects have been observed since the 1960s [\onlinecite{Miller1967, Dynes1978, Dynes1984, Lee1988}], 
and are often ascribed to electron-phonon ($e$-ph) coupling-induced relaxation and recombination of quasiparticles into the superconducting condensate of Cooper pairs \cite{Langenberg1975, Dynes1978}.

However, there are a number of additional effects which lead to a similar broadening in $\mathrm dI/\mathrm dV$: thin film effects~\cite{Rothwarf1967}, 
thermal fluctuations~\cite{Brihuega2011}, 
anisotropic energy gaps~\cite{Blackford1969}, electron-electron scattering~\cite{Reizer198998}, 
and inelastic scattering in dirty superconductors~\cite{Bose2009}.
In modern STS literature, there is often confusion about the origin of lifetime broadening \cite{Nishio2006, Bergeal2006}, 
motivating a thorough investigation.

In this \jw{work}, we present a STS study of the quasiparticle lifetime effects in thin lanthanum films grown on W(110) as a model system. In an earlier publication \cite{Loeptien2014}, we presented our findings about $\Delta_\mathrm{La}$ and the superconducting transition temperature $T_c$ for La films between the bulk limit and the thin film limit. \jwc{Here, we focus on} the broadening parameter $\Gamma_\mathrm{La}$ [\onlinecite{Dynes1978}]. We observe surprisingly large values of \jwb{$\Gamma_\mathrm{La}$}, a monotonous increase with the inverse film thickness $1/d$, and a decrease with $\Delta_\mathrm{La}$. These findings are evaluated taking into account the different physical phenomena mentioned above. We conclude that the dominant effect is an efficient relaxation of the quasiparticles at the interface to the normal metal substrate.

La films were prepared \textit{in situ} as described in Ref. [\onlinecite{Loeptien2014}] and studied in a commercially available UHV STM \cite{SPECS} at a base temperature of $T = 1.2$~K and at an elevated temperature of $T = 4.3$~K using normal metal tungsten and superconducting Nb tips \cite{Wiebe2004}. During the annealing, La forms flat-top (0001) islands in the dhcp phase in a Stranski-Krastanov growth, with a wetting layer of one monolayer in between the islands. Islands with thicknesses $d$ in the range between $d = 2.5$~nm and $d =140$~nm were grown~\cite{Loeptien2014}, which covers a wide range compared to the superconductor coherence length of lanthanum (36~nm) [\onlinecite{Pan1980}]. For the STS spectra, the differential conductivity $\mathrm dI/\mathrm dV$ was recorded using standard lock-in technique, adding a modulation voltage $V_\mathrm{mod} = 0.04$ to 0.07 mV (RMS value, modulation frequency $\nu_\mathrm{mod} = 0.93$~kHz) to the bias voltage $V$, stabilizing the tip 
 at a current of $I_\mathrm{stab} = 100$ to 150~pA and a voltage of $V_\mathrm{stab} = -6$~mV, opening the feedback and ramping $V$.

\begin{figure}[htbp]
	\centering
		\includegraphics[width=1\columnwidth]{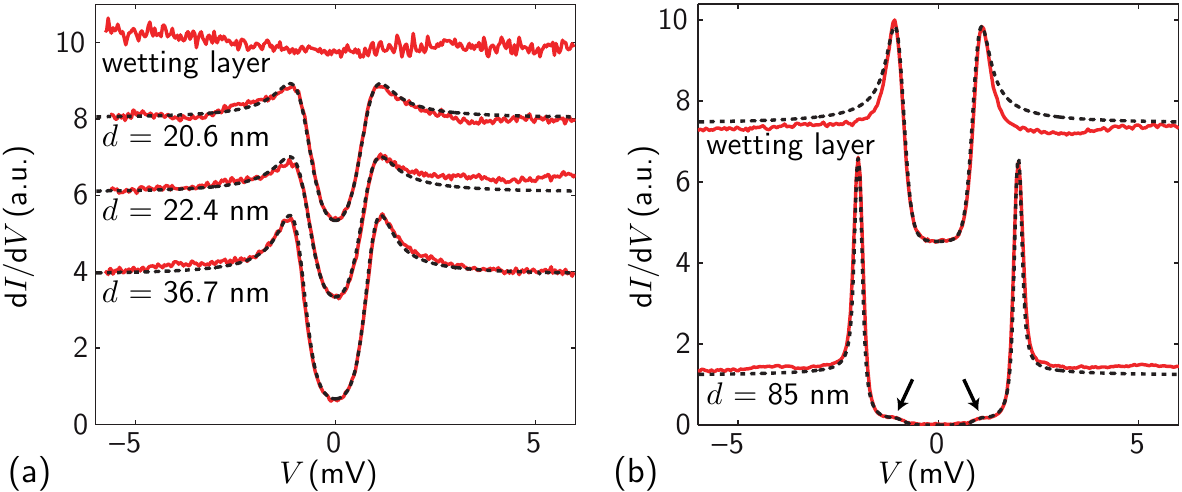}
		\caption{
		STS spectra (red solid lines) were taken on La islands with thicknesses $d$ as indicated and on the \jwb{wetting layer} to characterize the tip DOS, using (a) normal metal and (b) superconducting tips.
		Fitted calculations (dotted curves) are in excellent agreement with the experimental data and yield $\Delta_\mathrm{La}(T,d)$ and $\Gamma_\mathrm{La}(T,d)$.
		(a)~Experimental parameters: \mbox{$V_\mathrm{mod} = 0.06$~mV}, $T = 1.23$~K.	
		Model parameters: \mbox{$V_\mathrm{mod,eff} = 0.10$~mV}, \mbox{$\Delta_\mathrm{La} = 0.79$}, $0.80$, $0.89$~meV, $\Gamma_\mathrm{La} = 0.25$, $0.25$, $0.14$~meV (from top to bottom).	
		(b)~Experimental parameters: $V_\mathrm{mod} = 0.07$~mV, $T = 1.14$~K. 
		Model parameters: $V_\mathrm{mod,eff} = V_\mathrm{mod}$, $\Delta_\mathrm{tip} = 0.94$~meV, $\Gamma_\mathrm{tip} = 0.01$~meV, $\Delta_\mathrm{La} = 1.05$~meV, $\Gamma_\mathrm{La} = 0.07$~meV. 
		The curves are vertically shifted for better visibility \peter{by (a) 2~a.u. (b) 4.5~a.u.}		 
		Arrows: see text. (Reproduced from Figure 2 in [\onlinecite{Loeptien2014}])} 
	\label{fig:Figure_1}
\end{figure}

\begin{figure}[htbp]
	\centering
		\includegraphics[width=1\columnwidth]{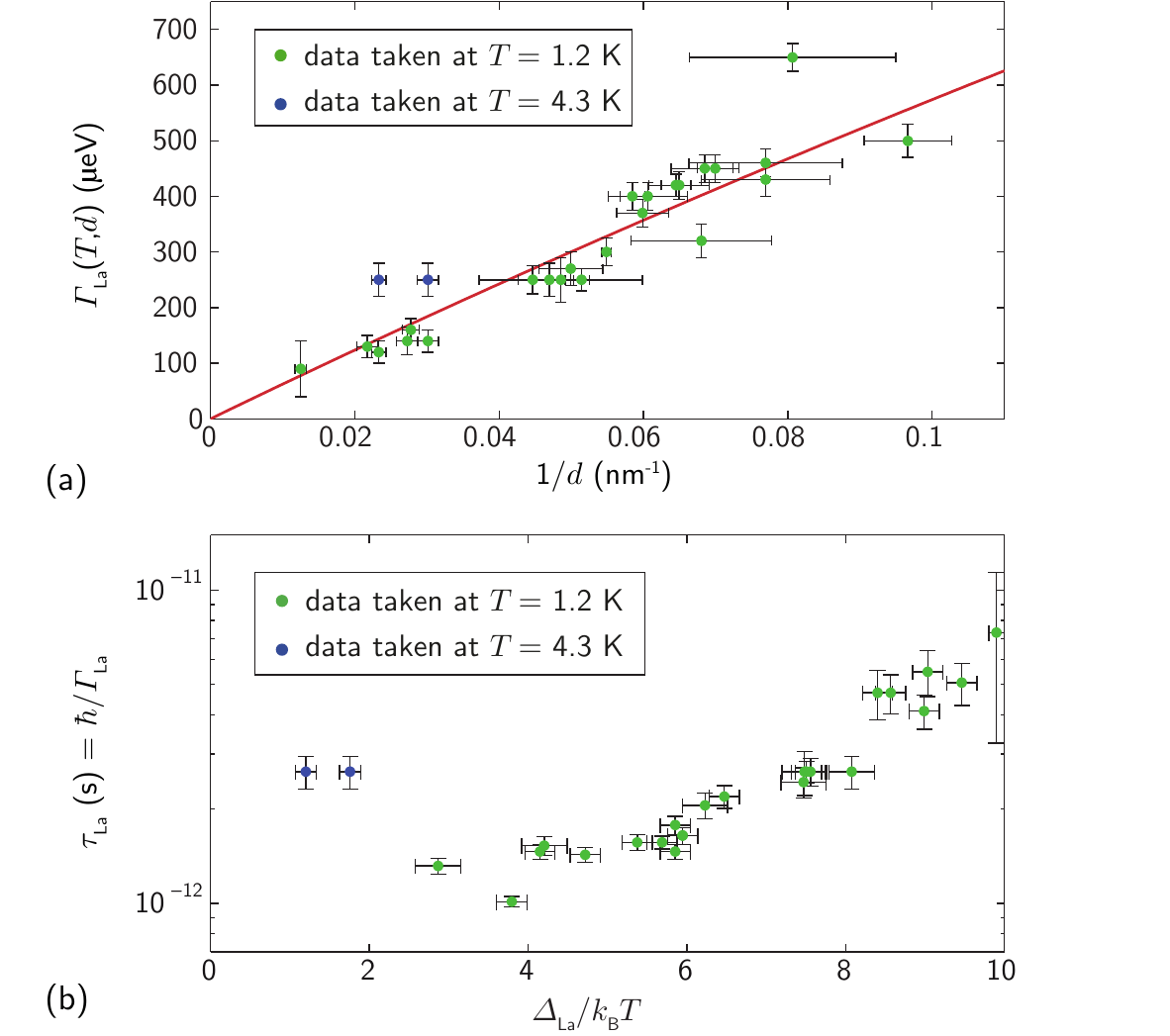}
		\caption{
		(a)~Dependence of $\Gamma_\mathrm{La}(T,d)$ on the inverse film thickness $1/d$. The fit (red line) is discussed in the text.
		(b)~Experimentally determined quasiparticle lifetime $\tau_\mathrm{La}= \hbar/\Gamma_\mathrm{La}(T,d)$ plotted logarithmically vs. $\Delta_\mathrm{La}(T,d)/k_\mathrm BT$.
		Green and blue data points indicate measurements taken at 1.2~K and 4.3~K, respectively. The error bars are due to uncertainties in the measured film thickness and in the fit parameters.
				}
	\label{fig:Figure_2}
\end{figure}

STS on La islands \jwb{using normal conducting [Fig.~\ref{fig:Figure_1}~(a)] or superconducting [Fig.~\ref{fig:Figure_1}~(b)] tips} reveals symmetric gaps with a width of $2\Delta_{\rm{La}}$ \jwb{or $2(\Delta_{\rm{La}}+\Delta_{\rm{tip}})$, respectively,} around the Fermi level $E_{\rm{F}}$ due to the superconductivity of the probed islands. Note, that by the use of the superconducting tip the energy resolution is strongly enhanced with respect to the thermal limit \cite{Ji2008, Franke2011}. The facts that the coherence peaks are rather wide and that there is considerable zero-bias conductance indicate strong broadening effects. In order to quantify the superconductivity and broadening via $\Delta_{\rm{La}}$ and $\Gamma_{\rm{La}}$, respectively,
the experimental curves were each fitted with a numerically simulated $\mathrm dI/\mathrm dV$, in analogy to the working principle of the lock-in amplifier \cite{Wiebe2004},
\begin{equation}
	\frac{\mathrm dI}{\mathrm dV}(V)\propto \int_{-\pi/2}^{+\pi/2} \sin(\alpha)\, I\left(V+\sqrt{2}\,V_\mathrm{mod, eff}\,\sin(\alpha),T\right)\,\mathrm d\alpha.
	\label{dIdV_SC_2}
\end{equation}
In this formula, $I$ is the tunneling current, 
\begin{multline}
	I(V, T)\propto \int_{-\infty}^{+\infty} N_1(E)\,N_2(E+eV) \times \left[f(E + eV, T)-f(E, T)\right]\, \mathrm d E\ ,
	\label{IV_SC}
\end{multline}
$N_1(E)$ and $N_2(E)$ are the DOSs of the two electrodes, and $f(E, T)$ is the Fermi function. For the normal metal electrodes, we assume a constant DOS on the relevant energy scale. The superconducting electrodes are modeled by a Dynes-like DOS \cite{Dynes1978},
\begin{equation}
	N_\mathrm{sc}(E, \Gamma) \propto \Re\left(\frac{E - i\,\Gamma}{\sqrt{(E - i\,\Gamma)^2-\Delta^2}}\right)
	\label{BCS_DOS_1_gamma}
\end{equation}
with a lifetime broadening parameter $\Gamma$\jwb{, where $\Re$ indicates the real part}. The excellent fit quality (Fig.~\ref{fig:Figure_1}) permits an accurate determination of $\Delta$ and $\Gamma$ for both, tip and sample.  
While most STS studies on superconductors \cite{Bergeal2006, Nishio2006, Brihuega2011} 
fit $\mathrm dI/\mathrm dV$-curves with the derivative of Eq.~\ref{IV_SC}, the approach of Eq.~\ref{dIdV_SC_2} considers a finite modulation voltage,
which enables to explicitly take into account the experimental broadening induced by the temperature, the lock-in technique, and additional electronic noise via an effective modulation voltage $V_\mathrm{mod,eff}\geq V_\mathrm{mod}$. \jw{The latter has been determined for each data set by fitting spectra taken with a Nb tip on the wetting layer which have negligible values of $\Gamma$.} 
This allows us to accurately determine the \textit{intrinsic} broadening parameters \jwb{$\Gamma$}.
The resulting values of $\Gamma_{\rm{La}}$ [Fig.~\ref{fig:Figure_2}(a)] are of the same order of magnitude as $\Delta_{\rm{La}}$ [\onlinecite{Loeptien2014}], and vary between $0.1$ and $0.6$~meV depending on the thickness $d$ of the island with a monotonous increase as a function of $1/d$.

The almost perfect reproduction of the STS spectra in the gap region by the Dynes-DOS (Eq.~\ref{BCS_DOS_1_gamma}), particularly of the in-gap shoulders \jw{which stem from the overlap of the coherence peak of one electrode with the broadening-induced non-zero in-gap DOS of the other electrode} [see arrows in Fig.~\ref{fig:Figure_1}(b)], strongly suggest an unusually short quasiparticle lifetime $\tau$ as the origin of the large $\Gamma_\mathrm{La}$. Other effects, that lead to such a broadening, \peter{\textit{i.e.}} averaging over different values of $\Delta$ due to anisotropic \cite{Blackford1969} or multigap \cite{Rodrigo2004a} superconductors, lead to an order of magnitude smaller values. Moreover, these effects offer no explanation for the $\Gamma(d)$-dependence. The extracted $\Gamma_{\rm{La}}$ of the La islands are 10 to 60 times larger as compared to typical $\Gamma_{\rm{tip}}$ values found for the Nb tips [c.f. Fig.~\ref{fig:Figure_1}(b)], substantiating that the broadening stems from the sample exclusively.
Therefore, the measured $\Gamma_\mathrm{La}$ can be directly converted into the quasiparticle lifetime in the sample via $\tau_\mathrm{La} = \hbar/\Gamma_\mathrm{La}$. This lifetime is plotted logarithmically as a function of the experimentally determined $\Delta_\mathrm{La}(T,d)/k_\mathrm BT$ in Fig.~\ref{fig:Figure_2}(b), showing a monotonous increase.
In the remaining part of this article, the dominant mechanisms leading to the measured quasiparticle lifetime in La will be discussed. The extracted lifetimes are considerably smaller \jwb{than} the typical lifetimes found for other conventional bulk superconductors~\cite{Miller1967, Hu1977, Dynes1978, Brihuega2011}, but comparable to typical values that have been reported for nanosized superconductors \cite{Nishio2006, Bose2009, Brihuega2011}. 
These facts already suggest that the short lifetimes must be linked to the finite size or the interfaces of the superconducting La film. \jw{Note, that spectra taken for a larger voltage range do not show any indication of quantum-well states on the La films~\cite{Loeptien2014}, which can be therefore neglected unlike, e.g., in thin superconducting Pb films \cite{Eom2006}.}

\begin{figure}[htbp]
	\centering
		\includegraphics[width=1\columnwidth]{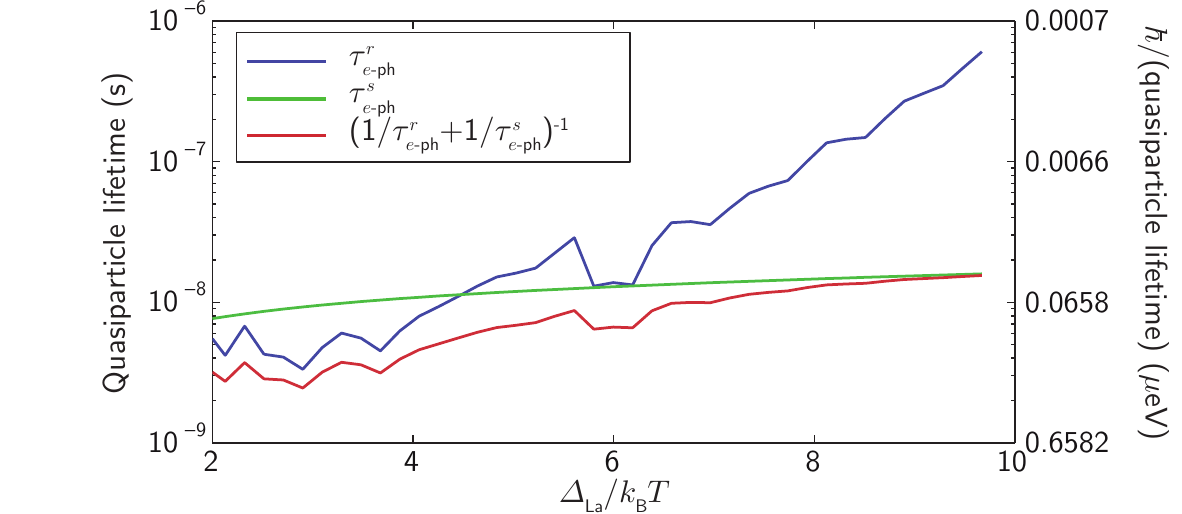}
	\caption{Calculated\cite{Kaplan1976} quasiparticle recombination time $\tau_{e-\mathrm{ph}}^r$, scattering time $\tau_{e-\mathrm{ph}}^s$\jwb{, and overall lifetime.} Input parameters: Theoretical $\alpha^2 F(E)$ of La [\onlinecite{Bagci2010}], $Z_1(0)=2.5$ [\onlinecite{Kaplan1976}], and \mbox{$T = 1.2$~K}. 
	}
	\label{fig:Figure_3}
\end{figure}

Let us first consider the usual electron-phonon ($e-\mathrm{ph}$) interaction-mediated processes of scattering with a rate $(\tau_{e-\mathrm{ph}}^s)^{-1}$ and quasiparticle recombination with a rate $(\tau_{e-\mathrm{ph}}^r)^{-1}$. 
In a theoretical model \cite{Kaplan1976}, 
these rates are calculated according to Eliashberg theory, using a theoretical\cite{Bagci2010} Eliashberg spectral function $\alpha^2 F(E)$ for La. This model assumes that the density of excess quasiparticles from tunneling, $n_\mathrm{tun}$, is much \jwb{smaller} than the density of thermally excited quasiparticles $n_\mathrm{th}$.
The resulting individual lifetimes and the overall $\tau = 1/(1/\tau^r_{e-\mathrm{ph}} + 1/\tau^s_{e-\mathrm{ph}})$ are \jwb{plotted in Fig.~\ref{fig:Figure_3} as a function of $\Delta_\mathrm{La}/k_\mathrm BT$ and reveal the same monotonous trend as the experimental data [c.f. Fig.~\ref{fig:Figure_2}~(b)]. However, the}
calculated lifetimes are 3--4 orders of magnitude too long in comparison to the measured lifetimes. Clearly, electron-phonon-mediated relaxation and recombination by thermally excited quasiparticles in the superconducting film alone can not explain the observed unusually short lifetimes.
The following additional effects which have been reported to change the lifetime can be neglected for this particular sample system:

(i) In thin superconducting films, excited quasiparticles may diffuse into the substrate where they become unavailable for recombination, or phonons can be trapped in the superconductor by backscattering at the interface if there is a large acoustic mismatch between the materials~\cite{Rothwarf1967}. 
Both these effects lead to even larger quasiparticle lifetimes in conflict with the experimental result.

(ii) Thermal fluctuations were shown~\cite{Brihuega2011} to evoke a broadening similar to Eq.~\ref{BCS_DOS_1_gamma}, with an exponentially increasing $\Gamma_\mathrm{La}$ for $T \rightarrow T_c$ [\onlinecite{Brihuega2011}]. Taking into account the measured weak temperature dependency of $\Gamma_\mathrm{La}$ [green vs. blue points in Fig.~\ref{fig:Figure_2}(a)], thermal fluctuations must have a negligible effect at the measurement temperature of $T = 1.2$~K.

(iii) Electron-electron interaction in a clean film should only play a role in case of a high Debye temperature \cite{Kaplan1976}, which is not the case in La [\onlinecite{Pan1980}].

(iv) Previous tunneling studies \cite{Dynes1984, Lee1988, Bose2009} ascribed high values of $\Gamma$ to inelastic scattering in the dirty limit, as the \jwb{electron-electron} scattering rate increases in disordered superconducting films \cite{Lee1988, DevBelitz1991}.
However, the expected $T^{9/2}$- or $T^{7/2}$-dependent \jwb{electron-electron} scattering rate is in contrast to the experimentally observed weak temperature dependence of $\Gamma_{\rm{La}}$ [green vs. blue points in Fig.~\ref{fig:Figure_2}(a)]. Moreover, an implausibly high disorder parameter \cite{DevBelitz1991} would be needed to describe the large values of $\Gamma_\mathrm{La}$. Altogether, this excludes disorder-enhanced scattering as a relevant mechanism.

(v) Finally, we have to consider whether the constant injection from the tunneling current $I$ generates a large density $n_\mathrm{tun}$ of excess quasiparticles. If $n_\mathrm{tun}\gg n_\mathrm{th}$, \jwb{which is called the over-injection regime, $n_\mathrm{tun}$} can lead to a considerable decrease in $\tau$, as already discussed for planar tunnel junction experiments.~\cite{Rothwarf1967} 
In order to estimate, whether our experiments were conducted in this regime, we consider that $n_\mathrm{tun}$, which is injected into an effective sample volume $\mathcal{V}$, annihilates with the theoretical recombination rate $\tau$ (Fig.~\ref{fig:Figure_3}) via $\mathrm d n_\mathrm{tun}/{\mathrm dt} = - n_\mathrm{tun}/\tau + I/(e\mathcal{V})$.~\cite{Rothwarf1967} Assuming the steady state condition $\mathrm d n_\mathrm{tun}/{\mathrm dt} = 0$, $n_\mathrm{tun}$ can be estimated from $\mathcal{V}$. Then, $n_\mathrm{tun}$ has to be compared to $n_\mathrm{th}$ as derived from the normal state DOS, which is assumed to be constantly $N(0)$ around $E_\mathrm F$,~\cite{Bagci2010} via an integration over an energy interval $[0,\Delta_\mathrm{La}]$, leading to $n_\mathrm{th} \approx \Delta_\mathrm{La} \, N(0) \, f(\Delta_\mathrm{La}, T)$. For a rough estimation of the effective volume $\mathcal{V}$, we consider the mean free path of the normal electrons extrapolated to $T = 0$~K of $l\approx30$~nm~[Ref.~\onlinecite{Soto2005}], and calculate $\mathcal{V}\approx l^3$. Together with $T=1.2$~K, $\Delta_\mathrm{La} = 0.8$~meV, and $I = 100$~pA, this results in $n_\mathrm{tun}/n_\mathrm{th} \approx 5$. Therefore, over-injection alone cannot explain the 3--4 orders of magnitude difference between the measured [Fig.~\ref{fig:Figure_2}~(b)] and calculated (Fig.~\ref{fig:Figure_3}) lifetimes.

Altogether, we conclude, that there must be an additional effect related to the reduced dimensionality in \jwc{thin} film geometry of the superconductor, which leads to the observed strong lifetime broadening. As already suggested by the large mean free path $l$, which is on the order of the thickness range of the films ($d = 2.5$~nm to $d =140$ \jwc{nm}), the quasiparticles \jwb{can} pass the superconducting film without effective relaxations, reaching the interface to the normal conducting substrate. \jwb{There}, efficient inelastic scattering processes at the interface \jwb{lead to} the strong lifetime reduction. In this scenario, the quasiparticle relaxation rate is mainly determined by the diffusion time to the interface, $\tau_\mathrm{diff}(d)$. It adds up with the relaxation time at the interface, $\tau_\mathrm{interface}$, to $\tau_\mathrm{full} = \tau_\mathrm{interface} + \tau_\mathrm{diff}(d)$. Only $\tau_\mathrm{diff}(d)$ depends on the film thickness $d$, simply by $1/\tau_\mathrm{diff}(d) = v_\mathrm{drift}/d$, introducing a thickness-independent drift velocity $v_\mathrm{drift}$. A fit of the experimental data $\Gamma_\mathrm{La}(d)$ to $\hbar/\tau_\mathrm{full}(d)$ [red line in Fig.~\ref{fig:Figure_2}(a)] yields a drift velocity $v_\mathrm{drift} = 1.5\times 10^{3}$~m/s\jwb{, which equals 1\% of the Fermi velocity \cite{Pan1980}.}
The required inelastic scattering rate of the interface as determined from the fit is $1/\tau_\mathrm{interface} = 1.6\times10^{12}\ \mathrm{s}^{-1}$.
Indeed, inelastic scattering rates of normal metal/superconductor interfaces in tunneling experiments were found to be as large as $1/\tau_\mathrm{interface}^\mathrm{lit} \approx 1/(47\ \mathrm{fs})$~[\onlinecite{Grajcar1995}], supporting the importance of the described interface relaxation of quasiparticles in order to explain the observed short lifetimes.

In summary, we determined an exceptionally short quasiparticle lifetime in thin superconducting La films grown on W(110). Considering all effects that have been reported in the literature, we find that only an efficient relaxation mechanism which is present at the superconductor-metal interface can explain the strong lifetime broadening, and leads to the observed dependence of the quasiparticle lifetime on the film thickness. 
Therefore, the short lifetime is not an intrinsic property of the superconducting material itself, and should be considered in future STS studies on thin-film superconductors.

We acknowledge financial support from the ERC Advanced Grant ``ASTONISH'', and from the DFG via Graduiertenkolleg 1286. 
A.A.K. acknowledges Project No. KH324/1-1 from the Emmy-Noether-Program of the DFG.
We thank A.~Eich for fruitful discussions.

\bibliography{library}

\newpage

\newpage

\newpage
\end{document}